\documentclass[12pt]{iopart}

\usepackage{iopams}
\usepackage{graphicx}

\begin{document}

\title{Algebraic equations for the exceptional eigenspectrum of the generalised Rabi model}

\author{Zi-Min Li$^{1}$ and Murray T. Batchelor$^{1,2,3}$}

\address{$^{1}$Centre for Modern Physics, Chongqing University, Chongqing 400044, China}

\address{$^{2}$Department of Theoretical Physics, 
Research School of Physics and Engineering, Australian National University, Canberra, ACT 0200, Australia}

\address{$^{3}$Mathematical Sciences Institute, Australian National University, Canberra ACT 0200, Australia}

\ead{batchelor@cqu.edu.cn}

\begin{abstract}
We obtain the exceptional part of the eigenspectrum of the generalised Rabi model, 
also known as the driven Rabi model, 
in terms of the roots of a set of algebraic equations.
This approach provides a product form for the wavefunction components 
and allows an explicit connection with recent results obtained for the wavefunction 
in terms of truncated confluent Heun functions.  Other approaches are also compared. 
For particular parameter values the exceptional part of the 
eigenspectrum consists of doubly degenerate crossing points.
We give a proof for the number of roots of the constraint polynomials and discuss the number of crossing points. 

\end{abstract}

\section{Introduction}

Despite it's simplicity, the generalised Rabi model has been solved only 
recently \cite{Braak,Chen,Braak2,Heun2,others2}.
The Rabi model \cite{Rabi} describes the simplest matter-field interaction, 
namely between a two-level atom and a single-mode bosonic field. 
It is thus a fundamental textbook model in quantum optics \cite{book}.
The generalised Rabi model has hamiltonian
\begin{equation}
H=\omega \, a^{\dagger}a + g \, \sigma_x(a^{\dagger}+a)+ \Delta \, 
\sigma_z+\epsilon\,\sigma_x \,, \label{ham}
\end{equation}
where $\sigma_x$ and $\sigma_z$ are Pauli matrices for a two-level system with level splitting $\Delta$.
The single-mode bosonic field is described by the creation and destruction operators $a^\dagger$  and $a$ with 
$[a, a^\dagger] = 1$ and frequency $\omega$.
The interaction between the two systems is via the coupling $g$. 
The Rabi model has $Z_2$ symmetry (parity) which is broken by the addition of the term $\epsilon \, \sigma_x$ 
in the generalised version of the model. 
This additional term allows tunnelling between the two atomic states. 
The generalised Rabi model (\ref{ham}) 
is also referred to as the driven Rabi model \cite{Larson} and is 
relevant to the description of various hybrid mechanical systems \cite{Heun2,hybrid}. 
Although having an analytic solution, both the Rabi and generalised Rabi models do not appear to be integrable in general in the 
Yang-Baxter sense \cite{BZ}.
However, the existence of monodromy matrices in terms of Painlev\'e V has now been reported \cite{Queiroz}.

The generalised Rabi model (\ref{ham}) has been solved in two ways: (i) 
by mapping the problem to the Bargmann space of analytic functions \cite{Braak,Braak2}, and 
(ii) by using the Bogoliubov operator method \cite{Chen}. 
Using the former approach explicit expressions have been obtained \cite{Heun2,others2} for the wavefunction 
in terms of confluent Heun functions \cite{heunbook}.\footnote{The connection with 
confluent Heun functions was made earlier for the Rabi model \cite{others0,Zhong,others}.}
Of particular relevance here is the fact that the energy spectrum of the generalised Rabi model, 
although possessing no parity symmetry, 
still includes both regular and exceptional parts. 
The eigenspectrum can be determined from the analytical solution. 
The exceptional parts, known as Juddian isolated exact solutions \cite{Judd}, 
can be systematically found from the conditions under which the confluent 
Heun functions are terminated as finite polynomials \cite{Heun2}. 
Our interest here is with this exceptional part of the eigenspectrum, which we obtain using a different approach.

Indeed the exceptional part of the Rabi model eigenspectrum 
has been obtained using a number of different (though ultimately related) 
approaches.\footnote{See, e.g., Refs \cite{Braak,Reik,Kus,Kus2,Koc,Zhang,Braak3,plet,WY}. 
A different generalised Rabi model is considered in Ref.~\cite{plet} with both rotating and counter-rotating terms, i.e., 
interpolating between the Jaynes-Cummings and Rabi models.}
Each approach results in the simple eigenvalue expression of a shifted oscillator, 
however with the system parameters satisfying a constraint which becomes  
increasingly complicated for higher energy levels.
Of most relevance here is an approach which derives a 
set of Bethe-like algebraic equations whose solutions define the constraint among the system parameters \cite{Zhang,plet}. 
We apply this approach in Section 2 to obtain the exceptional part of the eigenspectrum of the generalised Rabi model (\ref{ham}), 
allowing an explicit connection  
with the results obtained for the wavefunction in terms of truncated confluent Heun functions \cite{Heun2}.
The approach used here provides a simple product form for the wavefunction components in terms of the algebraic roots. 
We conclude by discussing the relation between the various approaches in Section 3.

\section{Results}

In the Bargmann realisation \cite{Schw}
\begin{equation}
a^\dagger \to z ,  \quad a \to \frac{d}{dz}
\end{equation}
the hamiltonian (\ref{ham}) reads
\begin{equation}
H=\omega \, z \frac{d}{dz} + g \, \sigma_x \left(z+\frac{d}{dz} \right)+ \Delta \, 
\sigma_z+\epsilon\,\sigma_x  \,.
\label{hamB}
\end{equation}
Following, e.g., Ref.~\cite{Heun2}, in terms of the two-component wavefunction  
\begin{equation}
\psi(z) = \left( \begin{array}{c}
\psi_+(z) \\
\psi_-(z) \end{array} \right)
\end{equation}
the Schr\"odinger equation $H \psi = E \psi$ gives rise to a pair of coupled equations for $\psi_+(z)$ and $\psi_-(z)$, namely
\begin{eqnarray}
(\omega z +g) \frac{{d\psi_+}}{{dz}} + (g z + \epsilon- E) \psi_+ + \Delta \psi_- &=& 0 \,,\\
(\omega z -g) \frac{{d\psi_-}}{{dz}} - (g z + \epsilon+ E) \psi_- + \Delta \psi_+ &=& 0 \,.
\end{eqnarray}

Two sets of solutions for $\psi_+(z)$ and $\psi_-(z)$ can be obtained. 
For the first set, the substitution $\psi^1_{\pm} (z)= {\mathrm e}^{-gz/\omega} \phi^1_\pm(z)$ leads to the coupled equations
\begin{eqnarray}
\left[(\omega z +g) \frac{{d}}{{dz}} - \left(\frac{g^2}{\omega} + E - \epsilon\right) \right] \phi^1_+(z)  = - \Delta \phi^1_-(z) \,,\label{phi}\\
\left[(\omega z -g) \frac{{d}}{{dz}} - \left(2gz -\frac{g^2}{\omega} + E + \epsilon\right) \right] \phi^1_-(z)  = - \Delta \phi^1_+(z)  \,.
\end{eqnarray}
Eliminating $\phi^1_-(z)$ gives the second order differential equation
\begin{eqnarray}
(\omega z - g) (\omega z +g) \frac{{d^2\phi^1_+(z)}}{{d^2z}} & \nonumber\\
+ \left[ - 2g\omega z^2 + (\omega^2 - 2 g^2 - 2 E \omega) z + 
\frac{g}{\omega} (2g^2 - \omega^2 - 2 \epsilon\omega)\right] \frac{{d\phi^1_+(z)}}{{dz}} & \nonumber\\
+\left[  2g\left(  \frac{g^2}{\omega}  + E -\epsilon\right) z + E^2 - \Delta^2 - \epsilon^2 + \frac{2\epsilon g^2}{\omega} - 
\frac{g^4}{\omega^2}     \right] \phi_+(z) = 0 \,.
\label{ode}
\end{eqnarray}

This is a case of the general second order differential equation considered by Zhang \cite{Zhang2}. 
Applying the result of Theorem 1.1 therein \cite{Zhang2}  gives the wavefunction component in the factorised form
\begin{equation}
\psi^1_+ (z)= {\mathrm e}^{-gz/\omega} \prod_{i=1}^N (z-z_i) \label{fac}
\end{equation}
where the roots $z_i$ satisfy the set of algebraic equations (details are given in Appendix A)
\begin{eqnarray}
\sum_{j \ne i}^N \frac{2\omega}{z_i - z_j} &=& \frac{2\omega^2 g z_i^2 + \left(2N \omega - \omega + 2\epsilon \right) \omega^2 z_i + 
\omega^2 g + 2  \epsilon \omega g - 2g^3}{\omega^2 z_i^2 - g^2}  \nonumber \\
&=& \frac{N \omega^2 + 2 \epsilon \omega}{\omega z_i - g} + \frac{N \omega^2 -\omega^2}{\omega z_i + g} + 2g
\label{alg1}
\end{eqnarray}
for $i=1,\ldots,N$. 
The system parameters obey the constraint 
\begin{equation}
\Delta^2 + 2 N g^2  + 2 \omega g \sum_{i=1}^N z_i = 0 \,. \label{con1}
\end{equation}
The energy of these states is given by 
\begin{equation}
E= N \omega  - \frac{g^2}{\omega} + \epsilon \,.  \label{energy+}
\end{equation}
The corresponding wavefunction component $\psi^1_- (z)$ can be determined using the result (\ref{fac}) and equation (\ref{phi}).
For $\epsilon=0$ the algebraic equations (\ref{alg1}) reduce to those obtained by the same approach \cite{Zhang}. 
The energy expression (\ref{energy+}) has been given in Ref.~\cite{Heun2}, where it follows as a condition for the 
general solution given in terms of the confluent Heun functions to truncate to a polynomial with $N$ terms.

Another set of solutions follow from the 
substitution $\psi^2_{\pm} (z)= {\mathrm e}^{gz/\omega} \phi^2_\pm(z)$,  leading to the coupled equations
\begin{eqnarray}
\left[(\omega z +g) \frac{{d}}{{dz}} + \left(2gz +\frac{g^2}{\omega} - E + \epsilon\right) \right] \phi^2_+(z)  = - \Delta \phi^2_-(z) \,, \\
\left[(\omega z -g) \frac{{d}}{{dz}} - \left(\frac{g^2}{\omega} + E + \epsilon\right) \right] \phi^2_-(z)  = - \Delta \phi^2_+(z) \,. \label{phi2}
\end{eqnarray}
Proceeding as above, these equations can be solved for the wavefunction components in the form 
\begin{equation}
\psi^2_- (z)= {\mathrm e}^{gz/\omega} \prod_{i=1}^N -(z-z_i) \label{fac2}
\end{equation}
where the roots $z_i$ satisfy the algebraic equations
\begin{eqnarray}
\sum_{j \ne i}^N \frac{2\omega}{z_i - z_j} &=& \frac{-2\omega^2 g z_i^2 + \left(2N \omega - \omega - 2\epsilon \right) \omega^2 z_i - 
\omega^2 g + 2 \epsilon \omega g + 2g^3 }{\omega^2 z_i^2 - g^2} \nonumber\\
&=& \frac{N \omega^2 -\omega^2}{\omega z_i - g}  + \frac{N \omega^2 - 2 \epsilon \omega}{\omega z_i + g} - 2g
\label{alg2}
\end{eqnarray}
for $i=1,\ldots,N$. 
The system parameters now obey the constraint 
\begin{equation}
\Delta^2 + 2 N g^2  - 2 \omega g \sum_{i=1}^N z_i = 0 \,,  \label{con2}
\end{equation}
with energy  
\begin{equation}
E= N \omega  - \frac{g^2}{\omega} - \epsilon \,.  \label{energy-}
\end{equation}
The corresponding wavefunction component $\psi^2_+ (z) = {\mathrm e}^{gz/\omega} \phi^2_+(z)$ 
follows from the result (\ref{fac2}) and equation (\ref{phi2}).
The energy expression (\ref{energy-}) has also been given in Ref.~\cite{Heun2}, again following from the condition 
for truncation of the general solution given in terms of the confluent Heun functions. 
This other set of solutions was not considered for $\epsilon=0$ \cite{Zhang}.
The resemblance of algebraic equations of this type with Richardson BCS equations of Gaudin type has been noted \cite{plet}. 

There is clearly a symmetry between the two sets of solutions.
Namely the algebraic equations (\ref{alg1}) and (\ref{alg2}) are equivalent under the 
transformation $z_i \leftrightarrow -z_i, \epsilon \leftrightarrow - \epsilon$. 
This corresponds to the related symmetry $\psi_+^1(z,\epsilon) = \psi_-^2(-z,-\epsilon)$, $\psi_-^1(z,\epsilon) = \psi_+^2(-z,-\epsilon)$
in the wavefunction components.
This symmetry is further discussed in Ref. \cite{Heun2} and is well known in the $\epsilon=0$ case (see, e.g., Ref. \cite{Kus2}).
The $-$ sign has been inserted into equation (\ref{fac2}) to ensure this symmetry.

\subsection{Examples}

We now turn to some specific examples. First consider $N=1$. 
The energy is 
\begin{equation}
E=  \omega  - \frac{g^2}{\omega} + \epsilon 
\end{equation}
and the algebraic equations (\ref{alg1}) reduce to 
\begin{equation}
2 \omega^2  g z_1^2 + \left( \omega + 2 \epsilon \right) \omega^2 z_1 +  \omega^2 g + 2 \epsilon \omega g - 
2 g^3  = 0 \,.  
\end{equation}
The two solutions are
\begin{equation}
z_1 = - \frac{g}{\omega}, \quad \frac{2g^2 - \omega^2 - 2 \epsilon \omega}{2 \omega g} \,.  
\label{sol1}
\end{equation}
Substitution into the constraint (\ref{con1}) gives $\Delta^2=0$ and 
\begin{equation}
\Delta^2 + 4 g^2 = \omega^2 + 2 \epsilon \omega,  
\label{conN1}
\end{equation}
respectively.
The value $\Delta^2=0$ obtained from the first solution corresponds to the degenerate atomic limit, which we discuss further below.
The second solution in (\ref{sol1}) gives the wavefunction components 
\begin{eqnarray}
\psi_+^{1}(z)&=& {\mathrm e}^{-g z / \omega} 
\left( \frac{2 \omega g z + \omega^2 + 2 \epsilon \omega - 2 g^2 }{2\omega g} \right) \,, \\
\psi_-^{1}(z)&=& {\mathrm e}^{-g z/ \omega} \, \frac{\Delta}{2g} \, , 
\end{eqnarray}
where in the last equation, we made use of the simplifying constraint (\ref{conN1}).

On the other hand, equation (\ref{alg2}) becomes 
\begin{equation}
- 2 \omega^2  g z_1^2 + \left( \omega - 2 \epsilon \right) \omega^2 z_1 -  \omega^2 g + 2 \epsilon \omega g + 
2 g^3  = 0 \,.  
\end{equation}
The two solutions are
\begin{equation}
z_1 =  \frac{g}{\omega}, \quad \frac{-2g^2 + \omega^2 - 2 \epsilon \omega}{2 \omega g} \,.  
\end{equation}
Substitution into the constraint (\ref{con2}) gives $\Delta^2=0$ and $\Delta^2 + 4 g^2 = \omega^2 - 2 \epsilon \omega$, respectively. 
The relevant energy and wavefunction components are 
\begin{equation}
E=  \omega  - \frac{g^2}{\omega} - \epsilon \,,
\end{equation}
\begin{eqnarray}
\psi_+^{2}(z) &=& {\mathrm e}^{g z / \omega} \, \frac{\Delta}{2g} ,\\
\psi_-^{2}(z) &=& {\mathrm e}^{g z/ \omega} \left( \frac{-2 \omega g z + \omega^2 - 2 \epsilon \omega - 2 g^2 }{2\omega g} \right) ,
\end{eqnarray}
The results for $N=1$ agree with those obtained from the truncation of the confluent Heun functions \cite{Heun2}, 
within a harmless renormalisation of the wavefunction components.

As a further check, consider $N=2$ for which equations (\ref{alg1}) are seen to give six sets of solutions. 
For the first set, $z_1=z_2 = -g/\omega$, the constraint relation  (\ref{con1}) gives $\Delta^2=0$. 
The solution $z_1=z_2 = g/\omega$ gives the unphysical constraint $\Delta^2 + 8g^2=0$.
As in dealing with Bethe Ansatz equations, the equations need to be solved numerically for finite sizes. 
For the simplest case $\epsilon=0$, the solutions with distinct roots have 
\begin{equation}
z_1 + z_2 =  \frac{-5 + 4\tilde{g}^2 \pm \sqrt{9 + 8\tilde{g}^2 + 16\tilde{g}^4}}{4\tilde{g}} \,, 
\end{equation}
with $\tilde{g}=g/\omega$.
Substitution into the constraint relation (\ref{con1}) and squaring gives the known result, namely 
$\Delta^4 + 12 \Delta^2 g^2 - 5 \Delta^2 \omega^2 + 32 g^4 
  - \,32 \omega^2 g^2   + 4 \omega^4 = 0.
$
Equations (\ref{alg2}) and (\ref{con2}) give the same constraint. 
The explicit wavefunction components for $N=2$ and $\epsilon \ne 0$ are
\begin{eqnarray}
\psi_+^{1}(z)&=& {\mathrm e}^{-g z / \omega} \, \left(z^2 + a_1 z + a_2^\epsilon\right) , \\
\psi_-^{1}(z)&=& {\mathrm e}^{-g z/ \omega} \, \left( b_1 z + b_2^\epsilon \right)  , \\
\psi_+^{2}(z) &=& {\mathrm e}^{g z / \omega} \, \left(-b_1 z + b_2^{-\epsilon} \right) ,\\
\psi_-^{2}(z) &=& {\mathrm e}^{g z/ \omega}  \, \left(z^2 - a_1 z + a_2^{-\epsilon}\right) ,
\end{eqnarray}
where
\begin{eqnarray}
 a_1&=& \frac{\Delta^2 + 4 g^2}{2 g \omega} , \qquad 
a_2^\epsilon= \frac{\Delta^4 + 8 \Delta^2 g^2 + 8 g^4 - \Delta^2 \omega^2 - 2 \Delta^2 \epsilon \omega }{8 g^2 \omega^2}  , \\
b_1 &=& \frac{\Delta}{2g} , \qquad
b_2^\epsilon = \frac{\Delta^4 + 6 \Delta^2 g^2  - \Delta^2 \omega^2 - 2 \Delta^2 \epsilon \omega }{4 \Delta g^2 \omega^2} .
\end{eqnarray}

\subsection{Degenerate atomic limit}

Some comments can be made about the degenerate atomic limit $\Delta=0$ for general $N$.
The degenerate solutions $z_i = - g/\omega$, for $i=1,\ldots,N$ satisfy the algebraic equations (\ref{alg1}), with $\Delta^2=0$ 
following from the constraint relation (\ref{con1}). 
The energy is given by (\ref{energy+}) with (\ref{fac}) giving the wavefunction component
\begin{equation}
\psi_+^{1}(z)= {\mathrm e}^{-g z / \omega} 
\left( z + \frac{g}{\omega} \right)^N . 
\end{equation}
This is precisely the solution obtained for the equivalent displaced harmonic oscillator in the Bargmann space \cite{Schw}.
The related solution \cite{Kus} similarly follows from equations (\ref{fac2})--(\ref{energy-}).

\section{Discussion}

It is interesting to compare the various approaches for deriving the exceptional part of the eigenspectrum. 
We have derived a set of algebraic equations (\ref{alg1}) for the exceptional part of the eigenspectrum of the generalised Rabi model (\ref{ham}) 
using a method  \cite{Zhang,Zhang2} akin to the functional or analytic Bethe Ansatz.
Although the energies have a simple form (\ref{energy+}), the constraint relations (\ref{con1}) and wavefunction components (\ref{fac}) 
are given in terms of the Bethe-like roots $z_j$. 
The constraint relations can be generated by a number of methods.
It is known, for example,  that the coefficients of the wavefunction components satisfy a system of $2N+1$ linear equations, with the constraint emerging 
as a condition for the determinant to vanish.
One can also determine a recurrence relation leading to the constraint relations (see, e.g., \cite{Koc}).

For the generalised Rabi model considered here, in terms of the series expansion coefficients $h_n$ 
for the confluent Heun function $\sum h_n x^n$, where $x=\frac{g-z}{2g}$, 
the recurrence relation is
\begin{equation}
A_n h_{n} = B_n  h_{n-1} + C_n h_{n-2}  \,, 
\label{relation} 
\end{equation}
with initial conditions $h_{-1}=0 $, $h_{0}= 1$.
The coefficients are given by
\begin{eqnarray}
A_n &=& n(-1  + n-N -2\epsilon/\omega) \, ,\\
B_n &=& (1-n+N)^2  - 4(n-1) g^2/\omega^2- \Delta^2/\omega^2 + 2(1-n+N)\epsilon/\omega \,, \\
C_n &=& 4 (-2 +n -N) g^2/\omega^2 \, .
\end{eqnarray}
This result follows from Ref.~\cite{Heun2} specified to the exceptional 
points.\footnote{Note that, taking $\omega=1$ and $\epsilon=0$ for simplicity, this recurrence relation can also be written in the form 
$(m+1)(m-N) h_{m+1} + \left[ \Delta^2 + 4g^2 m - (N-m)^2 \right] h_m + 4g^2 (N+1-m) h_{m-1}=0$, 
which differs from the recurrence relations given elsewhere, e.g., in Refs~\cite{Kus,Koc}. 
Presumabley this is because the coefficients in the recurrence relation change with the expansion variable, 
in this case either $x$ defined above or $z$.}
Indeed, the general three-term recurrence relation is central to the analytic solution 
of the generalised Rabi model \cite{Braak,Chen,Braak2,Heun2,others2}. 
The approach taken here effectively gives the factorisation of the truncated confluent Heun functions 
at the exceptional points.

The recurrence relation (\ref{relation}), and in particular the condition $C_{N+2}=0$,
 ensures that the infinite series expansion for the confluent Heun function
terminates with $h_{n}=0$ for $n > N$. 
The value $h_{N+1}=0$ determines the constraint relation for given $N$.
The first few polynomials obtained in this way are  
\begin{eqnarray}
&& \Delta^2 = 0 \,,\\
&& \Delta^2 \left( \Delta^2 + 4g^2 - \omega^2 - 2 \epsilon \omega \right) =0 \, , \label{N1}\\
&&  \Delta^2 \left( \Delta^4 + \Delta^2 (12 g^2 - 5 \omega^2 - 6 \epsilon \omega) 
+ 32 g^4  - 32 \epsilon g^2 \omega 
 + 8 \epsilon^2 \omega^2 \right.  \nonumber \\ 
&& \qquad - \,\left. 32  \omega^2 g^2 + 12 \epsilon \omega^3 + 4 \omega^4 \right) =0\,, \label{N2} \\
&& \Delta^2  (\Delta^6 + 2 \Delta^4 (12g^2 -7\omega^2 - 6\epsilon\omega) + 
\Delta^2 (49\omega^4 + 44 \epsilon^2 \omega^2 - 232 \omega^2 g^2  \nonumber\\
&&  \qquad + 176g^4 +16 \epsilon\omega (6\omega^2 - 11g^2)) - 12( -32g^6 + 24 \omega g^4 (2 \epsilon + 3 \omega) \nonumber\\
&& \qquad - 12 \omega^2 g^2 (2 \epsilon^2 + 5 \epsilon \omega + 3 \omega^2) + 4 \epsilon^3 \omega^3 \nonumber\\
&& \qquad + 12 \epsilon^2 \omega^4 + 11 \epsilon \omega^5 + 3 \omega^6 ) = 0 \,,
\end{eqnarray}
for $N=0,1,2,3$, respectively.
The $N=1$ result (\ref{N1}) is as given in (\ref{conN1}), 
with (\ref{N2}) the example given in Ref.~\cite{Heun2}. 
The constraint polynomials for given $N$ are generated readily enough via the recurrence relation. 
A similar recurrence relation can be written down corresponding to the solutions 
$\psi^2_{\pm} (z)= {\mathrm e}^{gz/\omega} \phi^2_\pm(z)$.
This results in the same constraint polynomials as given in the above examples, however with $\epsilon \leftrightarrow - \epsilon$.
In contrast the approach used here gives the closed form expressions (\ref{con1}) and (\ref{con2}), albeit in terms of the roots of the 
algebraic equations (\ref{alg1}) and (\ref{alg2}).
These equations remain to be explored.

\begin{figure}[t]
\begin{center}
\vskip 3mm
\includegraphics[width=0.7\columnwidth]{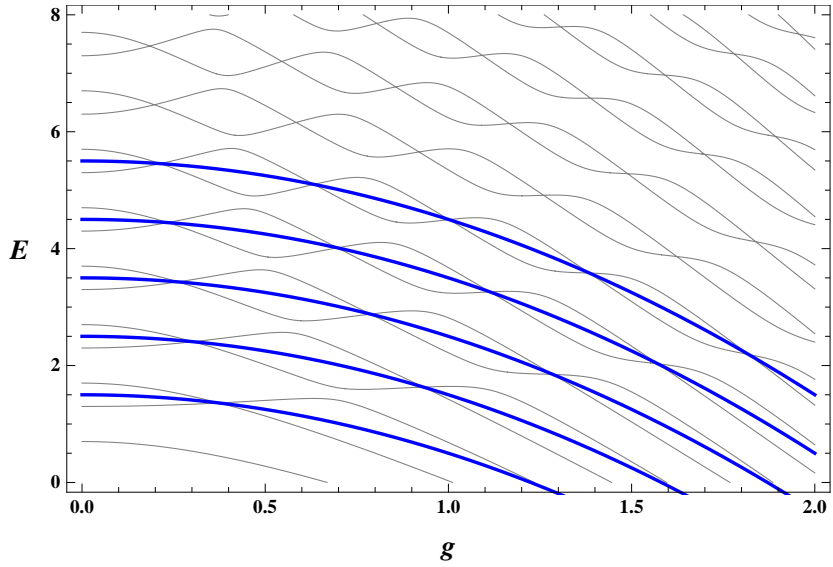}
\caption{The first few energy levels $E$ in the eigenspectrum of the generalised Rabi model as a function of the coupling $g$ 
denoted by grey (thin) lines. 
The parameter values are $\epsilon=\frac12 \omega$ with $\Delta=1.2$ and $\omega=1$. 
For this particular value of $\epsilon$ the exceptional part of the eigenspectrum consists of the doubly degenerate crossing points. 
The blue (thick) lines are the energy curves $E= N \omega - g^2/\omega + \epsilon$ for $N=1,\ldots,5$}
\label{fig1}
\end{center}
\end{figure}

\begin{figure}[t]
\begin{center}
\vskip 3mm
\includegraphics[width=0.7\columnwidth]{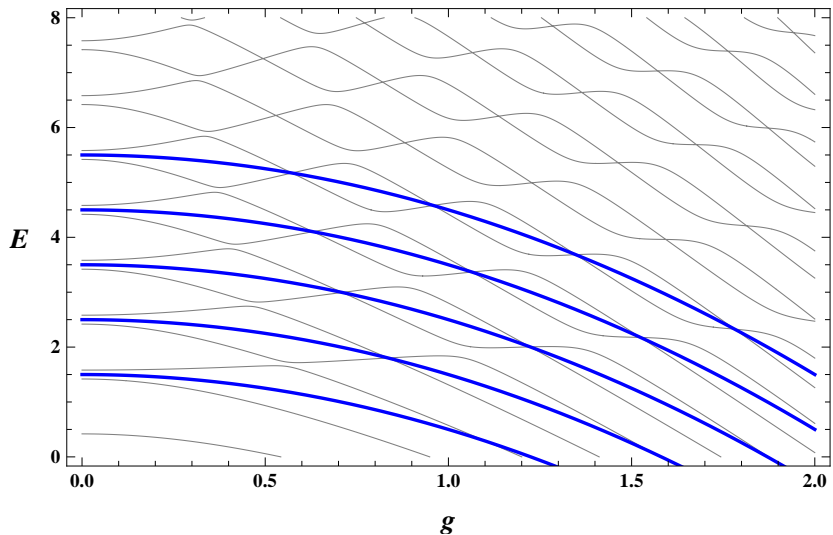}
\caption{The first few energy levels $E$ in the eigenspectrum of the generalised Rabi model as a function of the coupling $g$ 
denoted by grey (thin) lines. 
The parameter values are $\epsilon=\frac12 \omega$ with $\Delta=1.5$ and $\omega=1$. 
The blue (thick) lines are the energy curves $E= N \omega - g^2/\omega + \epsilon$ for $N=1,\ldots,5$.}
\label{fig2}
\end{center}
\end{figure}

\begin{figure}[t]
\begin{center}
\vskip 3mm
\includegraphics[width=0.7\columnwidth]{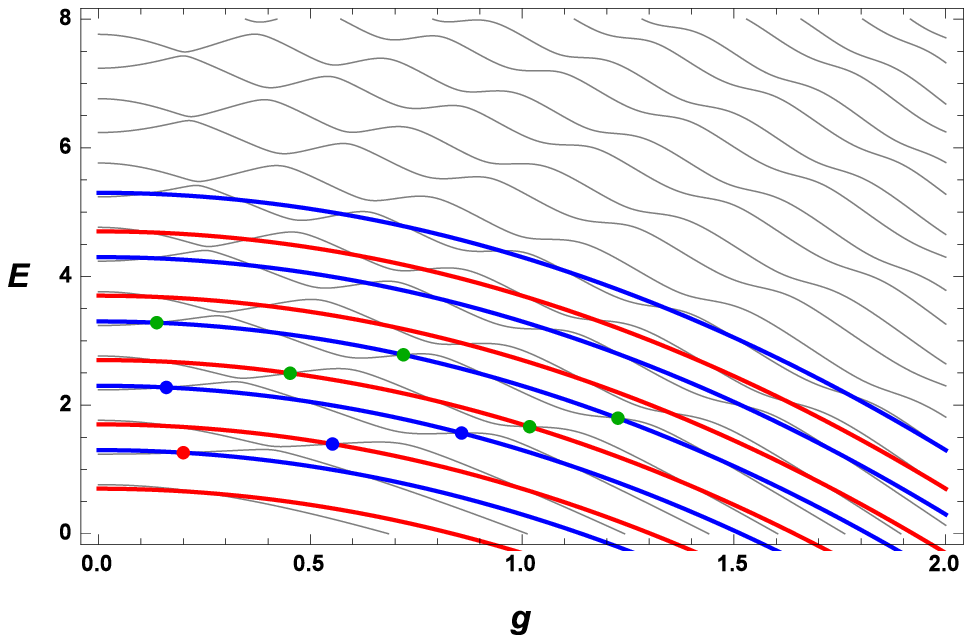}
\caption{The first few energy levels $E$ in the eigenspectrum of the generalised Rabi model as a function of the coupling $g$ 
denoted by grey (thin) lines. 
The parameter values are $\Delta=1.2$ and $\omega=1$ with now $\epsilon=0.3$. 
The blue (thick) lines are the energy curves $E= N \omega - g^2/\omega + \epsilon$ for $N=1,\ldots,5$. 
The red (thick) lines are the energy curves $E= N \omega - g^2/\omega - \epsilon$ for $N=1,\ldots,5$. 
For this value of $\epsilon$ there are no crossing points.
The first nine exceptional points in the eigenspectrum are indicated by red ($N=1$), blue ($N=2$) and green ($N=3$) circles. 
}
\label{fig3}
\end{center}
\end{figure}

\subsection{Degenerate crossing points}

It has been noted that when $\epsilon$ is an integer multiple of  $\omega/2$ the
exceptional eigenvalues considered here for the generalised Rabi model 
are crossing points in the eigenspectrum as a function of the coupling $g$ \cite{Braak,Heun2}.
For $\epsilon=0$ these are the well known Judd points, for which 
Kus \cite{Kus} provided a proof that for each $N$ there are $N$ crossings in the range $0 < \Delta < 1$. 
More generally Kus established that for $k < \Delta < k+1$ there are $N-k$ of them.
Figures~\ref{fig1} and \ref{fig2} show the first few levels in the eigenspectrum of the generalised Rabi model 
for $\epsilon=\frac12 \omega$ at different values of $\Delta$.
It is clear from Figure~\ref{fig1} that for the first few values of $N$ there are $N$ crossings for $\Delta=1.2$. 
We expect that this is indeed the case for all $N$ in the range $0 < \Delta < \sqrt{2}$.
Following Kus \cite{Kus} we are able to prove a theorem by induction in Appendix B giving the 
number of roots of the constraint polynomial and thus the number of exceptional points for given $N$ and $\epsilon$. 
Specifically, defining the function $Q_k(x)= (1/k!) P_k(x)$, where $P_N(x)=0$ is the constraint polynomial, 
we have established the following theorem.

\vskip 2mm

{\bf Theorem}. For $0 < \Delta/\omega < \sqrt{1+2 \epsilon/\omega}$, 
$Q_k(x)$ has exactly $k$ different, positive roots $a_1^{(k)}, a_2^{(k)}, \ldots, a_k^{(k)}$; moreover 
\begin{equation}
0 < a_1^{(k)} <   a_1^{(k-1)} <   a_2^{(k)} < a_2^{(k-1)} < \cdots < a_{k-1}^{(k-1)} < a_k^{(k)}\,, \label{rels}
\end{equation}
where $a_1^{(k-1)},\ldots, a_{k-1}^{(k-1)}$ denote the roots of $Q_{k-1}(x)$.

\vskip 2mm

More generally, as discussed in Appendix B, one can prove that there are $N-k$ roots of the constraint polynomial 
for $\Delta$ in the range 
\begin{equation}
\sqrt{k^2 + 2 k \epsilon/\omega} < \Delta/\omega < \sqrt{(k+1)^2 + 2 (k+1) \epsilon/\omega} \,.
\label{range}
\end{equation}
Indeed, in Figure~\ref{fig2} we see that for the first few values of $N$ there are $N-1$ crossings for $\Delta=1.5$ when $\epsilon=\frac12 \omega$. 
However, to prove the number of level crossings for $\epsilon=\frac12 \omega$ requires a further step. 
Figure~\ref{fig3} shows the first few exceptional points in the eigenspectrum at the typical value $\epsilon=0.3 \, \omega$ where there are 
no crossing points.
It was pointed out how these exceptional points merge to form doubly degenerate crossing points as 
$\epsilon \rightarrow \frac12 \omega$ \cite{Heun2}.
For the number of level crossings, 
it is necessary to prove that the two corresponding sets of roots of the constraint polynomials coincide for given $N$ 
when $\epsilon=\frac12 \omega$, see Appendix B.
This does not seem so straightforward to prove however, for general $N$.
Nevertheless, we are confident that there are indeed $N-k$ doubly degenerate crossing points in the range (\ref{range}). 
It would be fascinating, though seemingly unlikely, if such theorems could be proved via the algebraic equations obtained in this paper.

\ack
It is a pleasure to thank Professor Huan-Qiang Zhou for insightful discussions. 
We also thank Daniel Braak for suggesting to prove the number of crossings for $\epsilon = \frac12 \omega$ 
and the anonymous referees for a number of useful suggestions. 
MTB gratefully acknowledges support from Chongqing University and the 1000 Talents Program of China. 
This work is also supported by the Australian Research Council through grant DP130102839. 

\appendix

\section{Derivation of the algebraic equations}

The second order differential equation (\ref{ode}) is of the general form 
\begin{equation}
\left[ X(z) \frac{d^2}{dz^2} + Y(z) \frac{d}{dz} + Z(z) \right] S(z) =  0 \, ,
\label{Aode}
\end{equation}
where
\begin{equation}
X(z) = \sum_{k=0}^4 a_k x^k, \quad  Y(z) = \sum_{k=0}^3 b_k z^k, \quad Z(z)  =  \sum_{k=0}^2 c_k z^k \, .
\end{equation}
Comparing with equation (\ref{ode}), the nonzero coefficients are 
\begin{eqnarray}
a_0 &=& -g^2 \,, \quad a_2 = \omega^2 \,, \label{co1}\\
b_0 &=& \frac{2g^2}{\omega} - 2 \epsilon g - \omega g\,, \quad 
b_1 = \omega^2 - 2 g^2 - 2 E \epsilon \,, \quad
b_2 = - 2 \omega g\,, \\
c_0 &=& E^2 - \Delta^2 - \epsilon^2 + \frac{2\epsilon g^2}{\omega} - \frac{g^4}{\omega^2} \,, \quad
c_2 = \frac{2g^3}{\omega} + 2 E g - 2 \epsilon g \, . \label{co2}
\end{eqnarray}

Zhang's Theorem 1.1 \cite{Zhang2} states that (\ref{Aode}) has a degree $n$ polynomial solution 
\begin{equation}
S(z) = \prod_{i=1}^n (z-z_i)  \, , 
\end{equation}
with distinct roots $z_1, z_2, \ldots, z_n$.   
The values of the coefficients $c_0, c_1, c_2$ are given by 
\begin{eqnarray}
c_2 &=& - n(n-1) a_4 - n b_3 \,, \label{z1}\\
c_1 &=& -[2(n-1)a_4 + b_3] \sum_{i=1}^n z_i - n(n-1) a_3 - n b_2 \,, \\
c_0 &=& -[2(n-1)a_4 + b_3] \sum_{i=1}^n z_i^2 - 2 a_4 \sum_{i<j}^n z_i z_j \nonumber\\
&& -[2(n-1)a_3 + b_2] \sum_{i=1}^n z_i - n(n-1) a_2 - n b_1 \,. \label{z3}
\end{eqnarray}
The roots $z_1, z_2, \ldots, z_n$ satisfy the algebraic equations 
\begin{equation}
\sum_{j\ne i}^n \frac{2}{z_i-z_j} + \frac{b_3 z_i^3 + b_2 z_i^2 + b_1 z_i + b_0}{a_4 z_i^4 + a_3 z_i^3 + a_2 z_i^2 + a_1 z_i + a_0} = 0  \, , 
\label{z4}
\end{equation}
for $i = 1, 2, \ldots, n$.

Substituting the values (\ref{co1})-(\ref{co2}) into (\ref{z1})-(\ref{z4}) gives $0=0$, the energy expression (\ref{energy+}), the 
constraint (\ref{con1}) and the algebraic equations (\ref{alg1}), respectively. 
In a similar fashion we arrive at equations (\ref{alg2}), (\ref{con2}) and (\ref{energy-}).

\section{Proof for the number of exceptional points}

To prove the number of roots of the constraint polynomial it is convenient to generalise the recurrence relation obtained by 
Kus \cite{Kus} rather than the recurrence relation (\ref{relation}).
For nonzero $\epsilon$ we use the recurrence relation
\begin{eqnarray}
P_0 &=& 1 \,, \quad P_1 = 4g^2 + \Delta^2 - \omega^2 - 2 \epsilon \, \omega\,, \nonumber \\
P_k &=& \left[ k(2g)^2 + \Delta^2 - k^2 \omega^2 - 2k \epsilon \, \omega \right] P_{k-1} \nonumber \\ 
&& - k(k-1)(n-k+1) (2g)^2 \omega^2 P_{k-2}
 \,. \label{rec1}
\end{eqnarray}
The equation $P_k=0$ when $k=N$ defines the constraint polynomial, which can be written here in the form  
\begin{equation}
\left[ N(2g)^2 + \Delta^2 - N^2 \omega^2 - 2N \epsilon \, \omega \right] P_{N-1} - N(N-1)(2g)^2 \omega^2 P_{N-2} = 0 .
\end{equation}
We now fix the value of $N$ and set $x=(2g)^2$.
Thus
\begin{eqnarray}
Q_0(x) &=& 1 \,, \quad Q_1(x) = x- \alpha_1 \,, \nonumber \\
Q_k(x) &=& (x-\alpha_k) Q_{k-1}(x) - \beta_k \, x \, Q_{k-2}(x) 
 \,, \label{recq}
\end{eqnarray}
where $Q_k(x) = (1 / k!)  P_k(x)$ and 
\begin{eqnarray}
 \alpha_k &=& (k^2 \omega^2 + 2k \epsilon \, \omega - \Delta^2)/k, \label{alpha}\\
\beta_k &=&  (n - k +1) \omega^2 \,. 
\end{eqnarray}
Now, following Kus \cite{Kus}, we can prove the theorem stated in section 3.1.

{\sl Proof}. For $0 < \Delta/\omega < \sqrt{1+2 \epsilon/\omega}$ we have $\alpha_k > 0$ with always $\beta_k > 0$. 
From the definitions we also have $a_1^{(1)}=\alpha_1 > 0$ and $Q_2(x) = (x-\alpha_2) (x-a_1^{(1)}) - \beta_2 \, x $. 
Thus $Q_2(0) = \alpha_2 \, a_1^{(1)} > 0$, $Q_2(a_1^{(1)}) = - \beta_2 \, a_1^{(1)} < 0$ and $\mathrm{sgn} \, Q_2(\infty) = 1$, where 
\begin{equation*}
\mathrm{sgn} \, a =  \left\{ \begin{array}{rl}
         -1 & \quad \mbox{$a < 0$}\\
          0 & \quad \mbox{$a = 0$} \,\, .\\
          1 & \quad \mbox{$a > 0$} \end{array} \right.                     
\end{equation*}
These results and relations (\ref{rels}) prove that $Q_2(x) = (x-a_1^{(2)}) (x-a_2^{(2)})$ and $0 < a_1^{(2)} < a_1^{(1)} < a_2^{(2)}$. 

The general proof now proceeds by induction.
Assume that the theorem is valid for $\ell < k$, i.e., 
\begin{eqnarray}
Q_{k-1}(x) &=& (x - a_1^{(k-1)}) \cdots (x - a_{k-1}^{(k-1)})\,, \nonumber \\
Q_{k-2}(x) &=& (x - a_1^{(k-2)}) \cdots (x - a_{k-2}^{(k-2)})\,, 
\end{eqnarray}
and
\begin{equation}
0 < a_1^{(k-1)} <   a_1^{(k-2)} <   a_2^{(k-1)} < a_2^{(k-2)} < \cdots < a_{k-2}^{(k-2)} < a_{k-1}^{(k-1)}\,. \label{relsi}
\end{equation}
Then from (\ref{recq}) we have 
\begin{eqnarray}
 Q_k(x) &=& (x-\alpha_k) (x - a_1^{(k-1)}) \cdots (x - a_{k-1}^{(k-1)}) \nonumber \\
&& - \beta_k \, x \,  (x - a_1^{(k-2)}) \cdots (x - a_{k-2}^{(k-2)}) \,.
\end{eqnarray}
Thus 
\begin{eqnarray}
\quad \mathrm{sgn} \, Q_k(0) &=&  \mathrm{sgn} \, (-1)^k \, \alpha_k   \, a_1^{(k-1)} a_2^{(k-1)}  \cdots a_{k-1}^{(k-1)}  = (-1)^k \,, \nonumber \\
\mathrm{sgn} \, Q_k(a_i^{(k-1)}) &=& - \mathrm{sgn} \, 
\left( \beta_k \, a_i^{(k-1)} \, (a_i^{(k-1)}-a_1^{(k-2)}) \cdots (a_i^{(k-1)}-a_{i-1}^{(k-2)}) \right. \nonumber \\
&& \left. \times (a_i^{(k-1)}-a_i^{(k-2)}) \cdots (a_i^{(k-1)}-a_{k-2}^{(k-2)}) \right) = (-1)^{k-i} \,, \nonumber \\
\quad \mathrm{sgn} \, Q_k(\infty) &=&  1 \,.
\end{eqnarray}
This implies that 
\begin{equation}
Q_k(x) = (x-a_1^{(k)}) \cdots (x-a_k^{(2)}) \,,
\end{equation}
and $a_1^{(k-1)}, \ldots, a_1^{(k-1)}$ fulfill the inequalities (\ref{rels}).
The theorem is thus proved.

Again following Kus \cite{Kus} similar theorems can be proved for other values of $\Delta$.  
The ranges of $\Delta$ follow from the values of $k$ for which $\alpha_k$ defined in (\ref{alpha}) is no longer positive.
In this way one can prove that there are $N-k$ roots of the constraint relation in the range (\ref{range}). 
%
When $\epsilon=\frac12 \omega$ there are thus expected to be $N-k$ roots (exceptional points) in the range 
$\sqrt{k^2 + k} < \Delta/\omega < \sqrt{(k+1)^2 + k+1}$.
There are still expected to be $N-k$ points for given $N$ at $\Delta=\sqrt{k^2 + k}$.
However, for this special value we observe that at the left most edge of the energy plots two energy levels merge, rather than crossing, 
so technically they are not crossing points. 

To prove the double degeneracy at the crossing points we need to consider 
the other set of recurrence relations for nonzero $\epsilon$, which can be written as  
\begin{eqnarray}
P'_0 &=& 1 \,, \quad P'_1 = 4g^2 + \Delta^2 - \omega^2 - 2 \epsilon \, \omega\,, \nonumber \\
P'_k &=& \left[ k(2g)^2 + \Delta^2 - k^2 \omega^2 + 2k \epsilon \, \omega \right] P'_{k-1} \nonumber \\ 
&& - k(k-1)(n-k+1) (2g)^2 \omega^2 P'_{k-2}
 \,. 
\end{eqnarray}
The equation $P'_k=0$ when $k=N$ defines the constraint polynomial. 
We can repeat the above working to arrive at similar results for the number of roots and thus number of exceptional points.
Now one can prove that the function $Q'_N(x)=(1/k!)P'_N(x)$ has $N-k$ different positive roots for $\Delta$ in the range
\begin{equation}
\sqrt{k^2 - 2 k \epsilon/\omega} < \Delta /\omega < \sqrt{(k+1)^2 - 2 (k+1) \epsilon/\omega} \,.
\end{equation}

When $k=0$, $Q'_N(x)$ has $N$ roots for $0 < \Delta/\omega < \sqrt{1-2\epsilon/\omega}$.
We thus have that for $\sqrt{k^2 + 2 k \epsilon/\omega} < \Delta /\omega < \sqrt{(k+1)^2 - 2 (k+1) \epsilon/\omega}$ both 
$Q_N(x)$ and $Q'_N(x)$ have $N-k$ roots. 
In particular, for 
$\sqrt{(k+1)^2 - 2 (k+1) \epsilon/\omega} < \Delta /\omega < \sqrt{(k+1)^2 + 2 (k+1) \epsilon/\omega}$, 
$Q_N(x)$ has $N-k$ roots and $Q'_N(x)$ has $N-k-1$ roots.
Precisely at $\epsilon=\frac12 \omega$, the intervals $\left(\sqrt{k^2 + 2 k \epsilon/\omega},\sqrt{(k+1)^2 - 2 (k+1) \epsilon/\omega}\right)$ vanish.
This implies that for arbitrary $\Delta$, $Q_N(x)$ has $N-k$ roots and $Q'_N(x)$ has $N-k-1$ roots.
This is the situation seen, for example, in Figure 3.
We have been able to show numerically that the roots of the polynomials $Q_N(x)$ and $Q'_{N+1}(x)$ coincide 
for all values of $N$ when $\epsilon=\frac12 \omega$.\footnote{More generally the roots of $Q_N(x)$ and $Q'_{N+m}(x)$ coincide 
when $\epsilon=\frac12 m \omega$.}
The exceptional points are thus doubly degenerate at the crossing points.
However, we have not so far been able to prove this analytically.

\end{document}